\documentstyle[aps]{revtex} 

\newcommand\mathC{\mkern1mu\raise2.2pt\hbox{$\scriptscriptstyle|$}
		{\mkern-7mu\rm C}}						
\newcommand{\mathR}{{\rm I\! R}}                

\newcommand{\bra}[1]{\langle{#1}|}
\newcommand{\ket}[1]{|#1\rangle}
\newcommand{\braket}[2]{\bra{#1}{#2}\rangle} 
\newcommand{\ketbra}[2]{\ket{#1}\bra{#2}}

\begin{document}
 
\title{\hspace{12truecm} {\small Imperial/TP/96--97/64}\\
       \hspace{12truecm} {\small DAMTP-R97/55}\\[1cm]
	Continuous Time and Consistent Histories}

\author{C.J.~Isham\thanks{email: c.isham@ic.ac.uk}} 
\address{Blackett Laboratory, Imperial College, South Kensington,
	London SW7 2BZ, United Kingdom}
	
\author{N.~Linden\thanks{email: n.linden@newton.cam.ac.uk}} 
\address{D.A.M.T.P., University of Cambridge, Cambridge CB3 9EW,
	United Kingdom}

\author{K.~Savvidou\thanks{email: k.savvidou@ic.ac.uk}} 
\address{Blackett Laboratory, Imperial College, South Kensington,
	London SW7 2BZ, United Kingdom}

\author{S.~Schreckenberg\thanks{email: s.schreckenberg@ic.ac.uk}} 
\address{Blackett Laboratory, Imperial College, South Kensington,
	London SW7 2BZ, United Kingdom}

\maketitle 

\begin{center}
\medskip 			October, 1997
\end{center}

\begin{abstract} 
We discuss the use of histories labelled by a continuous time in the
approach to consistent-histories quantum theory in which
propositions about the history of the system are represented by
projection operators on a Hilbert space. This extends earlier work
by two of us \cite{IL95} where we showed how a continuous time
parameter leads to a history algebra that is isomorphic to the
canonical algebra of a quantum {\em field\/} theory.  We describe
how the appropriate representation of the history algebra may be
chosen by requiring the existence of projection operators that
represent propositions about time average of the energy.  We also
show that the history description of quantum mechanics contains an
operator corresponding to velocity that is quite distinct from the
momentum operator.  Finally, the discussion is extended to give a
preliminary account of quantum field theory in this approach to the
consistent histories formalism.

\end{abstract}
\pacs{ }

\section{Introduction}
The consistent-histories approach to quantum theory can be
formulated in several different ways. In the original scheme
\cite{Gri84,Omn88a,GH90b}, the crucial object is the decoherence 
function written as
\begin{equation} 
	d(\alpha,\beta)= {\rm tr}(\tilde C_\alpha^\dagger\rho 
		\tilde C_\beta) 							\label{Def:d}
\end{equation}
where $\rho$ is the initial density-matrix, and where the {\em class
operator\/} $\tilde C_\alpha$ is defined in terms of the standard
Schr\"odinger-picture projection operators $\alpha_{t_i}$ as
\begin{equation} 
	\tilde C_\alpha:=U(t_0,t_1)\alpha_{t_1} U(t_1,t_2)
	\alpha_{t_2}\ldots U(t_{n-1},t_n)\alpha_{t_n}U(t_n,t_0), 
													\label{Def:C_a} 
\end{equation}
where $U(t,t')=e^{-i(t-t')H/\hbar}$ is the unitary time-evolution
operator from time $t$ to $t'$. Each projection operator
$\alpha_{t_i}$ represents a proposition about the system at time
$t_i$, and the class operator $\tilde C_\alpha$ represents the
composite history proposition ``$\alpha_{t_1}$ is true at time
$t_1$, and then $\alpha_{t_2}$ is true at time $t_2$, and then
\ldots, and then $\alpha_{t_n}$ is true at time $t_n$''.

	As a product of (generically, non-commuting) projection
operators, the class-operator $\tilde C_\alpha$ is not itself a
projector. This difference between the representation of
propositions in standard quantum mechanics and in the history theory
is avoided in the alternative approach \cite{Isham94,IL94} to the
latter in which the history proposition ``$\alpha_{t_1}$ is true at
time $t_1$, and then $\alpha_{t_2}$ is true at time $t_2$, and then
\ldots, and then $\alpha_{t_n}$ is true at time $t_n$'' is
represented by the {\em tensor product\/}
$\alpha_{t_1}\otimes\alpha_{t_2}\otimes
\cdots\otimes\alpha_{t_n}$ which, unlike $\tilde C_\alpha$, {\em
is\/} a genuine projection operator. In this `history projection
operator' (HPO) scheme, the decoherence function can be written as
\begin{equation}
			d(\alpha,\beta)=\mbox{tr}_{{\cal V}\otimes{\cal V}}
					(\alpha\otimes\beta X)
\end{equation}
for a suitable operator $X$ (independent of $\alpha$ and $\beta$)
defined on ${\cal V}\otimes{\cal V}$ \cite{ILS94}. Here, $\cal V$
denotes the `history space' ${\cal H}_{t_1}\otimes {\cal
H}_{t_2}\otimes\cdots\otimes{\cal H}_{t_n}$ on which the
history-proposition projectors $\alpha$ and $\beta$ are defined,
where, for each $t_i$, ${\cal H}_{t_i}$ is a copy of the Hilbert
space ${\cal H}$ of the standard quantum theory.  The representation
of history propositions with projection operators has clear
advantages when discussing logical operations; in our case, quantum
temporal logic. It could also be applicable in quantum gravity
situations where there is no background concept of time and where,
therefore, the construction of a class operator is particularly
problematic: in this case, history propositions could still be
represented by projection operators, but on a Hilbert space that
does not arise as a temporal tensor product.

	The introduction of a {\em continuous\/} time clearly poses
difficulties for both approaches to the consistent history theory:
in the class-operator scheme one has to define continuous products
of projection operators; in the HPO approach, the problem is to
define a continuous {\em tensor\/} product of projection operators.

	In an earlier paper by two of us \cite{IL95}, the latter problem
was tackled by exploiting the well-known existence of continuous
tensor products of coherent states. However, several interesting
issues were sidestepped in the process. For example, the natural
history propositions in this scheme are represented by continuous
tensor products of projectors onto coherent states, and these do not
have a transparent physical interpretation. On the other hand, the
formalism as given was not well-equipped to handle the more
physically-motivated history propositions about continuous time
averages.

	In the present paper we take a fresh look at the question of
continuous time in the HPO formalism. As in our earlier work, the
starting point is the {\em history group\/}: a history-analogue of
the canonical group used in standard quantum mechanics.  The key
idea is that a unitary representation of the history group leads to
a self-adjoint representation of its Lie algebra, the spectral
projectors of which are to be interpreted as propositions about the
histories of the system. Thus we employ a history group whose
associated projection operators represent propositions about
continuous-time histories.  It transpires that the history algebra
for one-dimensional quantum mechanics is {\em infinite\/}
dimensional---in fact, it is isomorphic to the canonical commutation
algebra of a standard quantum {\em field\/} theory in one spatial
dimension.  This suggests that it might be profitable to study the
history theory using tools that are normally employed in quantum
field theory. This we shall do; in particular, we show that the
physically appropriate representation of the history algebra can be
selected by requiring the existence of operators that represent
propositions about the time-averaged values of the energy.  The Fock
space thus constructed is related to the notion of a continuous
tensor product as used in our earlier paper, thus establishing the
link with the idea of continuous temporal logic.  We also introduce
the continuous-time history analogue of the Heisenberg picture, and
we discuss the role of velocity in the history theory.
Finally, the discussion is extended to include the case of a free
relativistic quantum field.

	In what follows we have deliberately adopted a `physicist's
approach' to the analytical problems that arise in the theory; for
example, we frequently use unsmeared commutation relations, and
domains of operators are not discussed. This enables us to present
the essential physical ideas without getting lost in mathematical
detail. However, nothing of real importance is hidden thereby since
only Fock space representations of the history algebra are used, and
the full mathematical theory of these is well-known from normal
quantum field theory and poses no major problems.

\section{The History Space}
\subsection{The History Group}
We start by considering the HPO version of the quantum theory of a
particle moving on the real line $\mathR$. As explained above, the
history proposition ``$\alpha_{t_1}$ is true at time $t_1$, and then
$\alpha_{t_2}$ is true at time $t_2$, and then
\ldots, and then $\alpha_{t_n}$ is true at time $t_n$'' is
represented by the projection operator
$\alpha_{t_1}\otimes\alpha_{t_2}\otimes\cdots\otimes\alpha_{t_n}$ on
the $n$-fold tensor product ${\cal V}_n = {\cal H}_{t_1}\otimes{\cal
H}_{t_2}\otimes\cdots\otimes{\cal H}_{t_n}$ of $n$-copies of the
Hilbert-space $\cal H$ of the canonical theory.  Since $\cal H$
carries a representation of the Heisenberg-Weyl group with Lie
algebra 
\begin{equation} 
	{[\,}x,\, p\,] = i\hbar, 								\label{CCR} 
\end{equation}
the Hilbert space ${\cal V}_n$ carries a unitary representation of
the $n$-fold product group whose generators satisfy
\begin{eqnarray} 
	{[\,}x_k,\,x_m\,]&=& 0 					\label{discreteHWxx} \\
	{[\,}p_k,\,p_m\,]&=& 0 					\label{discreteHWpp} \\ 
	{[\,}x_k,\,p_m\,]&=& i\hbar\delta_{km} 	\label{discreteHWxp} 
\end{eqnarray} 
with $k,m = 1,2,\ldots,n$.  Thus the Hilbert space ${\cal V}_n$
carries a representation of the `history group' whose Lie algebra is
defined to be that of Eqs.\
(\ref{discreteHWxx})--(\ref{discreteHWxp}).  However, we can also
turn the argument around and {\em define\/} the history version of
$n$-time quantum mechanics by starting with Eqs.\
(\ref{discreteHWxx})--(\ref{discreteHWxp}). In this approach, ${\cal
V}_n$ arises as a representation space for Eqs.\
(\ref{discreteHWxx})--(\ref{discreteHWxp}), and tensor products
$\alpha_{t_1}\otimes\alpha_{t_2}\otimes\cdots\otimes\alpha_{t_n}$
that correspond to sequential histories about the values of position
or momentum (or linear combinations of them) are then elements of
the spectral representations of this Lie algebra.

	We shall employ this approach to discuss continuous-time
histories.  Thus, motivated by Eqs.\
(\ref{discreteHWxx})--(\ref{discreteHWxp}), we start with the
history-group whose Lie algebra (referred to in what follows as the
`canonical history algebra', or CHA for short) is \cite{IL95}
\begin{eqnarray}
{[\,}x_{t_1},\,x_{t_2}\,] &=&0 \label{ctsHWxx} \\
{[\,}p_{t_1},\,p_{t_2}\,] &=&0 \label{ctsHWpp} \\
{[\,}x_{t_1},\,p_{t_2}\,] &=&i\hbar\delta(t_1-t_2)\label{ctsHWxp}
\end{eqnarray} 
where $-\infty\leq t_1,\,t_2\leq\infty$.  Note that these operators
are in the {\em Schr\"odinger\/} picture: they must not
be confused with the Heisenberg-picture operators $x(t), p(t)$ of
normal quantum theory.

	An important observation is that Eqs.\
(\ref{ctsHWxx})--(\ref{ctsHWxp}) are mathematically the same as the
canonical commutation relations of a quantum {\em field \/} theory
in one space dimension: 
\begin{eqnarray}
{[\,}\phi(x_1),\phi(x_2)\,]&=&0				\label{1Dphiphi}\\
{[\,}\pi(x_1),\pi(x_2)\,]&=&0				\label{1Dpipi}\\
{[\,}\phi(x_1),\pi(x_2)\,]&=&i\hbar\delta(x_1-x_2).
											\label{1Dphipi}
\end{eqnarray}
This analogy will be exploited fully in the present paper. For
example, the following two issues arise immediately.  Firstly---to
be mathematically well-defined---equations of the type Eqs.\
(\ref{ctsHWxx})--(\ref{ctsHWxp}) must be smeared with test functions
to give
\begin{eqnarray}
	{[\,}x_f,\,x_g\,]&=&0                  	\label{SmearedCtsHWxx}  \\
	{[\,}p_f,\,p_g\,]&=&0                   \label{SmearedCtsHWpp}  \\
	{[\,}x_f,\,p_g\,]&=&i\hbar \int_{-\infty}^\infty f(t)g(t)\,dt,
								         	\label{SmearedCtsHWxp} 
\end{eqnarray}
which leads at once to the question of which class $\tau$ of test
functions to use. The minimal requirement for the right hand side of
Eq.\ (\ref{SmearedCtsHWxp}) to make sense is that $\tau$ must be a
linear subspace of the space $L^2(\mathR, dt)$ of square integrable
functions on $\mathR$. For the moment we shall leave $\tau$
unspecified beyond this.

	The second issue is concerned with finding the physically
appropriate representation of the CHA Eqs.\
(\ref{SmearedCtsHWxx})--(\ref{SmearedCtsHWxp}), bearing in mind that
infinitely many unitarily inequivalent representations are known to
exist in the analogous case of Eqs.\
(\ref{1Dphiphi})--(\ref{1Dphipi}).  Note that this problem does not
arise in standard quantum mechanics, or in the history version of
quantum mechanics with propositions defined at a finite number of
times, since---by the Stone-von Neumann theorem---there is a unique
representation of the corresponding algebra up to unitarily
equivalence.

	Of course, from the perspective of the history theory the
physically appropriate representation is expected to involve some
type of continuous tensor product; this was the path followed in our
earlier work \cite{IL95}. On the other hand, in standard quantum
field theory there is a folk lore, going back at least to a famous
paper by Araki \cite{Araki60}, to the effect that requiring the
Hamiltonian to exist as a proper self-adjoint operator is sufficient
to select a unique representation; for example, the representations
appropriate for a free boson field with different masses are
unitarily inequivalent.  In our case, this suggests that the
appropriate representation of the algebra Eqs.\
(\ref{SmearedCtsHWxx})--(\ref{SmearedCtsHWxp}) should be chosen by
requiring the existence of operators that represent history
propositions about (time-averaged) values of the energy. As we shall
see, this is indeed the case.

\subsection{The Hamiltonian Algebra}
We start with the ubiquitous example of the one-dimensional,
simple harmonic oscillator with Hamiltonian 
\begin{equation}
	H={p^2\over 2m}+{m\omega^2\over 2}x^2.
\end{equation}
The na{\"\i}ve idea behind the HPO theory is that to each time
$t$ there is associated a Hilbert space ${\cal H}_t$ that carries
propositions appropriate to that time (the `na{\"\i}vety' refers to
the fact that, in a {\em continuous\/} tensor product
$\otimes_{t\in\mathR}{\cal H}_t$, the individual Hilbert spaces
${\cal H}_t$ do not strictly exist as subspaces; this is related to
the need to smear operators).  Thus we expect to have a
one-parameter family of operators
\begin{equation} 
H_t:={p_t^2\over 2m}+{m\omega^2\over 2}x_t^2 \label{Def:Ht} 
\end{equation} 
that represent the energy at time $t$. 

	As it stands, the right hand side of Eq.\ (\ref{Def:Ht}) is not
well-defined, just as in normal canonical quantum field theory it is
not possible to define products of field operators at the same
spatial point.  However, the commutators of $H_t$ with the
generators of the CHA can be computed formally as
\begin{eqnarray}
&&{[\,}H_t,\,x_s\,]=-{i\hbar\over m}\delta(t-s)p_s 	\label{[Htxs]}\\
&&{[\,}H_t,\,p_s\,]=i\hbar m\omega^2\delta(t-s)x_s 	\label{[Htps]}\\
&&{[\,}H_t,\,H_s\,]=0 								\label{[HtHs]} 
\end{eqnarray} 
and are the continuous-time, history analogues of the familiar
result in standard quantum theory: 
\begin{eqnarray}
	{[\,}H,\,x\,]&=&-{i\hbar\over m }p 				\label{[Hx]}\\ 
	{[\,}H,\,p\,]&=&i\hbar m\omega^2 x.  			\label{[Hp]} 
\end{eqnarray}

	In standard quantum theory, the spectrum of the Hamiltonian
operator can be computed directly from the algebra of Eqs.\
(\ref{[Hx]})--(\ref{[Hp]}) augmented with the requirement that the
underlying representation of the canonical commutation relations
Eq.\ (\ref{CCR}) is irreducible. This suggests that we try to {\em
define\/} the history theory by requiring the existence of a family
of operators $H_t$ that satisfy the relations Eqs.\
(\ref{[Htxs]})--(\ref{[HtHs]}) and where the representation of the
canonical history algebra Eqs.\ (\ref{ctsHWxx})--(\ref{ctsHWxp}) is
irreducible.  More precisely, we augment the CHA with the algebra
(in semi-smeared form)
\begin{eqnarray}
	&&{[\,}H(\chi),\,x_t\,]=-{i\hbar\over m}\chi(t)p_t
												\label{[Hchixt]}\\
	&&{[\,}H(\chi),\,p_t\,]=i\hbar m\omega^2\chi(t)x_t 
												\label{[Hchipt]}\\
	&&{[\,}H(\chi_1),\,H(\chi_2)\,]=0 			\label{[Hchi1Hchi2]}
\end{eqnarray}
where $H(\chi)$ is the history energy-operator, time averaged with
the function $\chi$; heuristically,
$H(\chi)=\int_{-\infty}^{\infty}dt\, \chi(t) H_t$.

	It is useful to integrate these equations in the following
sense. {\em If\/} self-adjoint operators $H(\chi)$ exist satisfying
Eqs.\ (\ref{[Hchixt]})--(\ref{[Hchi1Hchi2]}), we can form the
unitary operators $e^{iH(\chi)/\hbar}$, and these satisfy
\begin{eqnarray} 
	&&e^{iH(\chi)/\hbar}\,x_t\,e^{-iH(\chi)/\hbar}
	=\cos[\omega\chi(t)]x_t +{1\over m\omega}\sin[\omega\chi(t)]p_t
								\label{UxtU}\\ 
	&&e^{iH(\chi)/\hbar}\,p_t\,e^{-iH(\chi)/\hbar}
		=-m\omega\sin[\omega\chi(t)]x_t+\cos[\omega\chi(t)]p_t.
														\label{UptU}
\end{eqnarray}								
However, it is clear that the right hand side of Eqs.\
({\ref{UxtU})--({\ref{UptU}) defines an {\em automorphism\/} of the
canonical history algebra Eqs.\ (\ref{ctsHWxx})--(\ref{ctsHWxp}).
Thus the task in hand can be rephrased as that of finding an
irreducible representation of the CHA in which these automorphisms
are unitarily implementable: the self-adjoint generators of the
corresponding unitary operators will then be the desired
time-averaged energy operators $H(\chi)$ [strictly speaking, weak
continuity is also necessary but this poses no additional problems
in the cases of interest here].

\subsection{The Fock Representation}
It is natural to contemplate the use of a Fock representation of the
CHA since this plays such a central role in the analogue of a free
quantum field in one spatial dimension. To this end, we start by
defining the `annihilation operator'
\begin{equation}
	b_t:=\sqrt{{m\omega\over2\hbar}}x_t+
			i\sqrt{{1\over 2m\omega\hbar}}p_t	\label{Def:bt} 
\end{equation}
in terms of which the CHA (\ref{ctsHWxx})--(\ref{ctsHWxp}) becomes
\begin{eqnarray}
	&&{[\,}b_t,\,b_s\,]=0							\label{[btbs]}\\
	&&{[\,}b_t,\,b_s^\dagger\,]=\delta(t-s).		\label{[btbsdag]} 
\end{eqnarray}
Note that 
\begin{equation}
	\hbar\omega b_t^\dagger b_s={1\over 2m}p_tp_s
		+{m\omega^2\over 2}x_tx_s-{\hbar\omega\over2}\delta(t-s) 
\end{equation}
which suggests that there exists an additively renormalised version
of the operator $H_t$ in Eq.\ (\ref{Def:Ht}) of the form
$\hbar\omega b_t^\dagger b_t$. In turn, this suggests strongly
that a Fock space based on Eq.\ (\ref{Def:bt}) should provide the
operators we seek.

	To make this explicit we recall that the bosonic Fock space
${\cal F}[{\cal H}]$ associated with a Hilbert space $\cal H$ is
defined as
\begin{equation}
	{\cal F}[{\cal H}]:=\mathC\oplus {\cal H}\oplus
			({\cal H}\otimes_S{\cal H})\oplus\cdots 
\end{equation}
where ${\cal H}\otimes_S{\cal H}$ denotes the symmetrised tensor
product of $\cal H$ with itself. Any unitary operator $U$ on the
`one-particle' space $\cal H$ gives a unitary operator $\Gamma(U)$
on ${\cal F}[{\cal H}]$ defined by 
\begin{equation}
	\Gamma(U):=
		1\oplus U\oplus (U\otimes U)\oplus\cdots	\label{Def:G(U)}
\end{equation} 
Furthermore, if $U=e^{iA}$ for some self-adjoint operator $A$ on
$\cal H$, then $\Gamma(U)=e^{id\Gamma(A)}$ where
\begin{equation}
	d\Gamma(A):=0\oplus A\oplus (A\otimes1+1\otimes A)\oplus\cdots.
												\label{Def:dGA}
\end{equation}

	The implications for us of these well-known constructions are as
follows. Consider the Fock space ${\cal F}[L^2(\mathR,dt)]$ that is
associated with the Hilbert space $L^2(\mathR,dt)$ via the
annihilation operator $b_t$ defined in Eq.\ (\ref{Def:bt}); {\em
i.e.,} the space built by acting with (suitably smeared) operators
$b_t^\dagger$ on the `vacuum state' $\ket{0}$ that satisfies
$b_t\ket{0}=0$ for all $t\in\mathR$.  The equations Eq.\
(\ref{UxtU})--(\ref{UptU}) show that, {\em if\/} it exists, the
operator $e^{iH(\chi)/\hbar}$ acts on the putative annihilation
operator $b_t$ as
\begin{equation} 
	e^{iH(\chi)/\hbar}\, b_t\, e^{-iH(\chi)/\hbar} =
		e^{-i\omega\chi(t)}b_t.  \label{UbtU} 
\end{equation} 
However, thought of as an action on $L^2(\mathR,dt)$, the operator
$U(\chi)$ defined by
\begin{equation}
		(U(\chi)\psi)(t):=e^{-i\omega\chi(t)}\psi(t) 
\end{equation}
is unitary for any measurable function $\chi$. Hence, using the
result mentioned above, it follows that in this particular Fock
representation of the CHA the automorphism on the right hand side of
Eq.\ (\ref{UbtU}) {\em is\/} unitarily implementable, and hence the
desired self-adjoint operators exist. Note that
$H(\chi)=\hbar\omega\, d\Gamma(\hat\chi)$, where the self-adjoint
operator $\hat\chi$ is defined on $L^2(\mathR,dt)$ as
\begin{equation}
	(\hat\chi\psi)(t):=\chi(t)\psi(t).
\end{equation}

	In summary, we have shown that the Fock representation of the
CHA Eqs.\ (\ref{ctsHWxx})--(\ref{ctsHWxp}) associated with the
annihilation operator $b_t$ of Eq.\ (\ref{Def:bt}) is such that
there exists a family of self-adjoint operators $H(\chi)$ for which
the algebra Eqs.\ (\ref{[Hchixt]})--(\ref{[Hchi1Hchi2]}) is
satisfied. This Fock space is the desired carrier of the history
propositions in our theory. Note that, in this case, the natural
choice for the test function space $\tau\subseteq L^2(\mathR,dt)$
used in Eqs.\ (\ref{SmearedCtsHWxx})--(\ref{SmearedCtsHWxp}) is
simply $L^2(\mathR,dt)$ itself.

	The position history-variable $x_t$ can be written in terms of
$b_t$ and $b_t^\dagger$ as
\begin{equation}
	x_t=\sqrt{{\hbar\over2m\omega}}\left(b_t+b_t^\dagger\right) 
\end{equation}
and has the correlation function
\begin{equation}
		\bra{0}x_t\,x_s\ket{0}={\hbar\over2m\omega}\delta(t-s).
								\label{<xtxs>}
\end{equation}
Thus the carrier space of our history theory is Gaussian white noise.

	Finally, we note that the formalism discussed above can be
extended to a wide class of Hamiltonian operators by employing the
types of argument used by axiomatic field theorists to construct
euclidean quantum field theories in one space plus one time
dimension \cite{Sim74}. In particular, the underlying Gaussian
stochastic process of our history theory of the simple harmonic
oscillator can be successfully perturbed by the addition to $H_t$ of
a wide class of polynomial functions of the configuration variable
$x_t$.

\subsection{The `$n$-particle' History Propositions} 
The Fock-space construction produces a natural collection of history
propositions: namely, those represented by the projection operators
onto what, in a normal quantum field theory, would be called the
`$n$-particle states'. To see what these correspond to physically in
our case we note first that a $\delta$-function normalised basis for
${\cal F}[L^2(\mathR,dt)]$ is given by the vectors $\ket{0}$,
$\ket{t_1}$, $\ket{t_1,t_2}$, \ldots where
$\ket{t_1}:=b_{t_1}^\dagger\ket{0}$, $\ket{t_1,t_2}:=b_{t_1}^\dagger
b_{t_2}^\dagger\ket{0}$, {\em etc\/} (of course, properly normalised
vectors are of the form $\ket{\phi}:=b_\phi^\dagger\ket{0}$ {\em
etc} for suitable smearing function $\phi$).  The physical meaning
of the projection operators of the form $\ketbra{t}{t}$ (or, more
rigorously, $\ketbra{\phi}{\phi}$), $\ketbra{t_1,t_2}{t_1,t_2}$,
{\em etc\/}, can be seen by studying the equations
\begin{eqnarray}
	&&H(\chi)\ket{0}=0 \\ &&H(\chi)\ket{t}=\hbar\omega\chi(t)\ket{t}
\\ &&H(\chi)\ket{t_1,t_2}=\hbar\omega[\chi(t_1)+\chi(t_2)]
\ket{t_1,t_2}
\end{eqnarray}
or, in totally unsmeared form,
\begin{eqnarray}
	&&H_t\ket{0}=0										\\
	&&H_t\ket{t_1}=\hbar\omega\delta(t-t_1)\ket{t_1}	\\
	&&H_t\ket{t_1,t_2}=\hbar\omega[\delta(t-t_1)+\delta(t-t_2)]
						\ket{t_1,t_2}.
\end{eqnarray}	

	It is clear from the above that, for example, the projector
$\ketbra{t_1,t_2}{t_1,t_2}$ represents the proposition that there is
a unit of energy $\hbar\omega$ concentrated at the time point $t_1$
and another unit concentrated at the time point $t_2$. Note that
$H(\chi)\ket{t,t}=2\hbar\omega\chi(t)\ket{t,t}$, and hence
$\ketbra{t,t}{t,t}$ represents the proposition that there are two
units of energy concentrated at the {\em single\/} time point $t$
(thus exploiting the bose-structure of the canonical history
algebra!).  This interpretation of projectors like
$\ketbra{t_1,t_2}{t_1,t_2}$ is substantiated by noting that
the time-averaged energy obtained by choosing the averaging
function $\chi$ to be $1$ acts on these vectors as
\begin{eqnarray}
	&&\int_{-\infty}^\infty ds\, H_s\ket{t}=\hbar\omega\ket{t}\\
	&&\int_{-\infty}^\infty ds\, H_s\ket{t_1,t_2}=
			2\hbar\omega\ket{t_1,t_2}
\end{eqnarray} 
and so on. This is the way in which the HPO account of the
simple harmonic oscillator recovers the integer-spaced energy
spectrum of standard quantum theory.

	Finally, we note in passing that
\begin{equation}
	{1\over\hbar\omega}\int_{-\infty}^\infty ds\,
	sH_s\,\ket{t_1,t_2,\ldots,t_n}=
			(t_1+t_2+\cdots+t_n)\ket{t_1,t_2,\ldots,t_n} 
\end{equation}
so that ${1\over\hbar\omega}\int_{-\infty}^\infty ds\, sH_s$ acts as
a `total time' or `center-of-time' operator.

\subsection{The Heisenberg Picture}
It is interesting to investigate the analogue of the Heisenberg
picture in our continuous-time HPO theory. In standard quantum
theory, the Heisenberg-picture version of an operator $A$ is defined
with respect to a time origin $t=0$ as
\begin{equation} 
	A_H(s):=e^{isH/\hbar}\,A\,e^{-isH/\hbar}. \label{Def:AH(t)} 
\end{equation} 
In particular, for the simple harmonic oscillator we have
\begin{eqnarray} 
x(s)&=&\cos[\omega s]x+{1\over m\omega}\sin[\omega s]p \\ 
p(s)&=&-m\omega\sin[\omega s]x+\cos[\omega s]p.
\end{eqnarray}
The Heisenberg-picture operator $x(s)$ satisfies the classical
equation of motion
\begin{equation}
		{d^2x(s)\over ds^2}+\omega^2 x(s)=0,		\label{EM} 
\end{equation}
and the commutator of these operators is 
\begin{equation}
	{[\,}x(s_1),\,x(s_2)\,]={i\hbar\over m\omega}\sin[\omega(s_1-s_2)]
							\label{[xs1xs2]}
\end{equation}
which, on using the equation of motion
\begin{equation}
	p:=m\left.{dx(s)\over ds}\right|_{s=0},		\label{Def:p} 
\end{equation}
reproduces the familiar canonical commutation relation Eq.\
(\ref{CCR}).

	In trying to repeat this construction for the history theory we
might be tempted to define the Heisenberg-picture analogue of, say,
$x_t$ as
\begin{equation}
	x_{H,t}(s):= e^{isH_t/\hbar}\,x_t\, e^{-isH_t/\hbar}.
\end{equation}
However, this expression is not well-defined since it corresponds to
choosing the test-function in Eq.\ (\ref{UxtU}) as
$\chi(t'):=s\delta(t-t')$, which leads to ill-defined products of
$\delta(t-t')$.

	What is naturally suggested instead is to define `time-averaged'
Heisenberg quantities
\begin{equation}
	x_{\kappa,t}:=e^{iH(\kappa)/\hbar}\,x_t\,e^{-iH(\kappa)/\hbar}
		=\cos[\omega\kappa(t)]x_t+
		{1\over m\omega}\sin[\omega\kappa(t)]p_t \label{Def:xkt}
\end{equation}
for suitable test functions $\kappa$. The analogue of the equation
of motion Eq.\ (\ref{EM}) is the functional differential
equation
\begin{equation} 
{\delta^2 x_{\kappa,t}\over\delta\kappa(s_1)\delta\kappa(s_2)} +
\delta(t-s_1)\delta(t-s_2)\omega^2x_{\kappa,t}=0,
\end{equation}
while the history analogue of Eq.\ (\ref{Def:p}) is
\begin{equation}
	\delta(t-s)p_t=m \left.{\delta
				x_{\kappa,t}\over\delta\kappa(s)}\right|_{\kappa=0}, 
\end{equation}
and the analogue of the `covariant commutator' Eq.\
(\ref{[xs1xs2]}) is
\begin{equation}
{[\,}x_{\kappa_1,t_1},\,x_{\kappa_2,t_2}\,]={i\hbar\over m\omega}
\delta(t_1-t_2)\sin[\omega(\kappa_1(t_1)-\kappa_2(t_2)]
\end{equation} 
which correctly reproduces the canonical history algebra.

	It is worth remarking that we could have proceeded in a slightly
different way by starting with a set of operators
$x_t(s)$ that satisfy a postulated history version of the covariant
commutator Eq.\ (\ref{[xs1xs2]}) 
\begin{equation}
	{[\,}x_{t_1}(s_1),\,x_{t_2}(s_2)] ={i\hbar\over m\omega}
	\delta(t_1-t_2)\sin[\omega(s_1-s_2)] 
\end{equation} 
and with the standard Schr\"odinger canonical history operators then
being {\em defined\/} as $x_t:=x_t(0)$ and $p_t:=m{dx_t(s)\over
ds}|_{s=0}$.  However, in fact, this is just a special case of the
first scheme with the test function $\kappa$ chosen to be
$\kappa(t):=s$ for all $t$; {\em i.e.,} the `Heisenberg picture'
operators $x_t(s)$ are generated by the time-averaged energy in the
form
\begin{equation}
			x_t(s):=e^{is\int dr\, H_r/\hbar}\,x_t\,e^{-is\int dr\,
				H_r/\hbar}.				\label{Def:xt(s)}
\end{equation}

	In our HPO formalism, the Heisenberg-picture operators---unlike
those in the Schr\"odinger picture---have no obvious direct physical
interpretation and their main use is likely to be mathematical.
Therefore, there is no {\em a priori\/} reason for rejecting the
simple time-averaged quantity in Eq.\ (\ref{Def:xt(s)}). What {\em
is\/} clear however is that---whichever version is used---{\em
two\/} different time labels appear: the `external' label $t$ that
specifies the time at which a proposition is asserted, and the
`internal' label $s$ that specifies the time parameter in the
Heisenberg picture associated with the copy of standard quantum
theory on the Hilbert space ${\cal H}_t$.

\subsection{The Relation With Continuous Temporal Logic} 
Relating the construction above to the idea of `continuous temporal
logic' involves showing that
\begin{equation}
	\otimes_{t\in\mathR}L^2_t(\mathR,dx)\simeq 	
				{\cal F}[L^2(\mathR,dt)]	\label{cont=FS} 
\end{equation}
where $L^2(\mathR,dx)$ is the Hilbert space of the standard quantum
theory of a particle moving in one dimension.  For completeness we
shall summarise here the discussion of this relation given in our
earlier paper \cite{IL95}.

	At a heuristic level, the inner product between two continuous
tensor product vectors $\ket{\otimes_{t\in\mathR}u_t}$ and
$\ket{\otimes_{t\in\mathR} v_t}$ is required to be 
\begin{equation}
	\braket{\otimes_{t\in\mathR}u_t}
		{\otimes_{t\in\mathR}v_t}_{\otimes_t{\cal H}_t}
	=\prod_{t\in\mathR}\braket{u_t}{v_t}_{{\cal H}_t}:= 
		\exp\int_{-\infty}^\infty dt\,
			\log\braket{u_t}{v_t}_{{\cal H}_t}.  \label{SPCTP}
\end{equation}
Since $\braket{u_t}{v_t}$ is a complex number the logarithm on the
right hand side of Eq.\ (\ref{SPCTP}) is generally not well-defined.
However, consider the special case of trying to construct the
continuous tensor product of copies of a {\em Fock\/} space ${\cal
F}[K]$ for some Hilbert space $K$. In particular, consider the {\em
coherent states\/} $\ket{\exp\phi}$, $\ket\phi\in K$, defined as
\begin{equation}
	\ket{\exp\phi}
		:=\oplus_{n=0}^\infty{1\over n!}(\otimes\ket\phi)^n
											\label{Def:expf} 
\end{equation}
where $(\otimes\ket\phi)^n$ denotes the tensor product of $\ket\phi$
with itself $n$ times (and with the convention that
$(\ket\phi)^0:=1$).  Then
\begin{equation}
	\braket{\exp\phi}{\exp\psi}_{{\cal F}[K]}=
			\exp\braket{\phi}{\psi}_{K}
\end{equation}
and hence, using Eq.\ (\ref{SPCTP}),  
\begin{equation}
\braket{\otimes_t\exp\phi_t}{\otimes_t
		\exp\psi_t}_{\otimes_t{\cal F}[K]_t}
	=\exp\int_{-\infty}^\infty dt\,\braket{\phi_t}{\psi_t}_{K_t}
												\label{dummy1}
\end{equation}
which {\em is\/} well-defined. But the exponent in the right hand
side of Eq.\ (\ref{dummy1}) is just the inner product in the direct
integral $\int^{\oplus}K_t\, dt$ of the Hilbert spaces $K_t$, and
hence we arrive at the basic isomorphism
\begin{eqnarray}
	\otimes_{t\in\mathR}{{\cal F}[K]_t}&\simeq& 
			{\cal F}[\int^\oplus K_t\,dt]	\label{bas-iso}	\\
	\otimes_t\ket{\exp\phi_t}&\mapsto&\ket{\exp\phi(\cdot)}.\nonumber 
\end{eqnarray}

	However, the single-time Hilbert space of our
theory---$L^2(\mathR,dx)$---can be written as the Fock space
for the one-dimensional Hilbert space $\mathC$ via the isomorphism
\begin{eqnarray} 
	{\cal F}[\mathC]&\simeq& L^2(\mathR,dx)\\
	\ket{\exp{z}}&\mapsto&\braket{x}{\exp z}:=
(		2\pi)^{-1/4}e^{zx-(1/2)z^2-(1/4)x^2} 
\end{eqnarray} where the right
hand side involves the familiar coherent states in $L^2(\mathR,dx)$.
Thus the isomorphism Eq.\ (\ref{bas-iso}) becomes \begin{equation}
\otimes_{t\in\mathR}L^2_t(\mathR,dx) \simeq {\cal F}[\int^\oplus
\mathC_t\,dt]. \label{bas-iso1} \end{equation} But the direct
integral $\int^\oplus \mathC_t\,dt$ is isomorphic to
$L^2(\mathR,dt)$ via the map that takes the parametrised family of
complex numbers $\lambda_t$ to the {\em function\/} $\lambda(\cdot)$
in $L^2(\mathR,dt)$ [{\em i.e.,} $\lambda(t):=\lambda_t$]. Hence we
arrive at the desired isomorphism in Eq.\ (\ref{cont=FS}).

\subsection{The Extension to Three Dimensions} 
The extension of the formalism above to a particle moving in three
spatial dimensions appears at first sight to be unproblematic. The
analogue of the history algebra Eqs.\
(\ref{ctsHWxx})--(\ref{ctsHWxp}) is
\begin{eqnarray}
	{[\,}x^i_{t_1},\,x^j_{t_2}\,] &=&0 \label{ctsHWxx3D} \\
{[\,}p^i_{t_1},\,p^j_{t_2}\,] &=&0 \label{ctsHWpp3D} \\
{[\,}x^i_{t_1},\,p^j_{t_2}\,] &=& i\hbar\delta^{ij}\delta(t_1-t_2)
									\label{ctsHWxp3D} 
\end{eqnarray} $i,j=1,2,3$; while the formal expression Eq.\
(\ref{Def:Ht}) for the energy at time $t$ becomes
\begin{equation}
	H_t:={\underline{p}_t\cdot\underline{p}_t\over 2m}+
{m\omega^2\over 2}\underline{x}_t\cdot\underline{x}_t.
\label{Def:Ht3D}
\end{equation}

	It is straightforward to generalise the discussion above to this
situation and, in particular, to find a Fock representation of Eqs.\
(\ref{ctsHWxx3D})--(\ref{ctsHWxp3D}) in which the rigorous analogues
of Eq.\ (\ref{Def:Ht3D}) exist as {\em bona fide\/} self-adjoint
operators. However, an interesting issue then arises that has no
analogue in one-dimensional quantum theory. Namely, we expect to
have angular-momentum operators whose formal expression is
\begin{equation} 
	L^i_t:=\epsilon^i{}_{jk}x^j_tp^k_t 			\label{Def:Lit}
\end{equation} 
and whose commutators can be computed heuristically as
\begin{equation}
{[\,}L^i_t,\,L^j_s\,]=i\hbar\epsilon^{ij}{}_k\delta(t-s)L^k_t.
												\label{[LitLjs]}
\end{equation}
Such operators $L^i_t$ can be constructed rigorously using, for
example, the method employed for the energy operators $H_t$: viz.,
compute the automorphisms of the canonical history algebra that are
formally induced by the angular-momentum operators and then see if
these automorphism can be unitarily implemented in the given Fock
representation. However, the interesting observation is that, even
if this can be done (which is the case, see below), this does not
guarantee {\em a priori\/} that the commutators in Eq.\
(\ref{[LitLjs]}) will be reproduced: in particular, it is necessary
to check directly if a $c$-number {\em central extension\/} is
present since we know from other branches of theoretical physics
that algebras of the type in Eq.\ (\ref{[LitLjs]}) are prone to such
anomalies.

An obvious technique for evaluating such a commutator would be to
define the angular momentum operators by point-splitting in the form
\begin{equation}
		L^i_{t,\epsilon} := i\hbar\epsilon^{i}{}_{jk}
			(b^j_{t})^\dagger b^k_{t+\epsilon}
\end{equation} 
so that the commutator in Eq.\ (\ref{[LitLjs]}) is the analogue of
an equal-time commutator in standard quantum field theory, and the
point-splitting is the analogue of spatial point splitting.  It is
then straightforward to compute the commutators of these point-split
operators and take the limit $\epsilon\rightarrow 0$. The
result is the anticipated algebra Eq.\ (\ref{[LitLjs]}).

	However, in standard quantum field theory it is known that the
limit of the commutator has to be considered at unequal times ({\em
i.e.}, using Heisenberg-picture operators), and that there is a
subtle relation between the two limits of the times becoming equal
and the spatial point splitting tending to zero\cite{Jackiw}.
Therefore, in order to calculate correctly the commutator in our
case it seems appropriate to consider the analogue of an unequal
time commutator, namely
\begin{equation}
	[L^i_{\chi,t,\epsilon},\,L^j_{0,s,\epsilon}] 
\end{equation} 
where
\begin{equation}
	L^i_{\chi,t,\epsilon}:= i\hbar\epsilon^{i}{}_{jk}(b^j_{\chi,t})^\dagger
		b^k_{\chi,t+\epsilon},
\end{equation}
and where
\begin{equation}
		 b^k_{\chi,t} := e^{iH(\chi)} b^k_t e^{-iH(\chi)} 
		 = e^{-i\omega\chi(t)} b^k_t
\end{equation}
is a time-averaged Heisenberg picture operator of the type defined
earlier.

	It is not difficult to show that
\begin{eqnarray}
		[L^i_{\chi,t,\epsilon},\,L^j_{0,s,\epsilon}] &&= 
	-\hbar^2 e^{i\omega(\chi(t)-\chi(t+\epsilon))}\nonumber\\ 
&&\times\big[\delta(t-s+\epsilon)\left((b^j_{t})^\dagger b^i_{t+2\epsilon} - 
\delta^{ij}(b^m_{t})^\dagger b^m_{t+2\epsilon} \right)\nonumber\\
&&\qquad-\delta(t-s-\epsilon) 
\left( (b^i_{t-\epsilon})^\dagger b^j_{t+\epsilon} -
\delta^{ij}(b^m_{t-\epsilon})^\dagger b^m_{t+\epsilon} \right)
\big]
\end{eqnarray}
and then, by evaluating the matrix element of the
commutator in the vacuum state, one sees that there is no central
extension in this case.  Furthermore, by considering the matrix
element of the commutator in general coherent states, one can check
that the limits of $\epsilon \rightarrow 0$ and $\chi\rightarrow 0$
are straightforward, and that as long as the test functions are
smooth, the angular momentum generators do indeed satisfy the
heuristic commutator Eq.\ (\ref{[LitLjs]}) in the limit.

\subsection{The Role of the Velocity Operator} 
The HPO approach to the consistent-histories theory has the striking
feature that, formally, there exists an operator that corresponds to
propositions about the {\em velocity\/} of the system: namely,
$\dot x_t:={d\over dt}x_t$. More rigorously, we can adopt the
procedure familiar from standard quantum field theory and define
\begin{equation} 
	\dot x_f:=-x_{\dot f} \label{Def:dot x}
\end{equation} 
which is meaningful provided that (i) the test-function $f$ is
differentiable; and (ii) $f$ `vanishes at infinity' so that the
implicit integration by parts used in Eq.\ (\ref{Def:dot x}) is
allowed; {\em i.e.}, heuristically, $x_f=\int_{-\infty}^\infty
dt\, x_t f_t$.

	The rigorous existence of $\dot x_t$ depends on the precise
choice of test-function space used in the smeared form of the CHA in
Eqs.\ (\ref{SmearedCtsHWxx})--(\ref{SmearedCtsHWxp}). In the
analogous situation in normal quantum field theory, the
test-functions are chosen so that the spatial derivatives of the
quantum field exist, this being necessary to define the Hamiltonian
operator. In our case, the situation is somewhat different since the
energy operator $H_t$ [see Eq.\ (\ref{Def:Ht})] does not depend on
$\dot x_t$ and hence there is no {\em a priori\/} requirement for
$\dot x_t$ to exist. However, what {\em is\/} clear from Eq.\
(\ref{ctsHWxx}) is that {\em if\/} $\dot x_t$ exists then
\begin{equation}
	{[\,}x_t,\,\dot x_s\,]=0				\label{[xtxdot]} 
\end{equation}
and hence our theory allows for history propositions that include
assertions about the position of the particle and its
velocity at the same time; in particular, the velocity $\dot x_t$
and momentum $p_t$ are not related. In this context it should be
emphasised once more that $x_t$, $t\in\mathR$, is a one-parameter
family of {\em Schr\"odinger\/}-picture operators---it is {\em
not\/} a Heisenberg-picture operator, and the equations of motion do
not enter at this level.

	The existence of a velocity operator that commutes with position
is a striking property of the HPO approach to consistent histories
and raises some intriguing questions. For example, a classic paper
by Park and Margenau \cite{PM68} contains an interesting discussion
of the uncertainty relations, including a claim that it {\em is\/}
possible to measure position and momentum simultaneously provided
the latter is defined using time-of-flight measurements. The
existence in our formalism of the vanishing commutator Eq.\
(\ref{[xtxdot]}) throws some new light on this old discussion. Also
relevant in this respect is Hartle's discussion of the operational
meaning of momentum in a history theory \cite{Har91b}. In
particular, he emphasises that an accurate measurement of momentum
requires a long time-of-flight, whereas---on the other hand---our
definition of velocity as the time-derivative of the history
variable $x_t$ clearly involves a vanishingly small time interval.
Presumably this is the operational difference between momentum and
velocity in the HPO approach to consistent histories.

	The potential existence of $\dot x_t$ also raises the
interesting possibility of defining a `velocity-extended' version of
the energy $H_t$ as
\begin{equation}
	H_t:={p_t^2\over 2m} +
	{m\over2}(\omega^2 x_t^2+\lambda \dot x_t^2)	\label{Def:Hvext} 
\end{equation}
for some real parameter $\lambda\geq 0$. Note that in the
one-dimensional quantum field theory analogue in which $x_t$ and
$p_t$ are replaced by $\phi(x)$ and $\pi(x)$ respectively, for an
appropriate choice of $\lambda$ the expression in Eq.\
(\ref{Def:Hvext}) becomes the usual Hamiltonian density $H(x)$ for a
massive scalar field (the $\lambda=0$ case of Eq.\ (\ref{Def:Ht})
then correponds to an {\em ultralocal\/} quantum field). Guided by
this observation, we can try to repeat our earlier analysis and look
for a representation of the canonical history algebra in which a
suitably smeared version of this new energy exists as a proper
operator. The quantum field theory analogue sugests that the
appropriate replacement for Eq.\ (\ref{Def:bt}) is
\begin{equation} 
	c_t:=\left({m\sqrt{\omega^2-\lambda D^2}\over 2\hbar} 
			\right)^{1/2} x_t +
	i\left({1\over2m\hbar\sqrt{\omega^2-\lambda D^2}} 
			\right)^{1/2} p_t						\label{Def:ct} 
\end{equation} 
where $D$ denotes the differential operator $d\over dt$. Note that
$-D^2$ is a positive semi-definite operator on $L^2(\mathR,dt)$, and
hence the square-root of $\omega^2-\lambda D^2$ is well-defined; of
course, to make all this rigorous a suitably-smeared form of Eq.\
(\ref{Def:ct}) should be used.

	We note that
\begin{equation}
	{[\,}c_t,\,c_s^\dagger\,]=\delta(t-s)
\end{equation}
and that, {\em if\/} it existed, the smeared form $H(\chi)$ of the
velocity-extended Hamiltonian would generate the CHA automorphism
\begin{equation} 
	e^{iH(\chi)/\hbar}\, c_t\, e^{-iH(\chi)/\hbar} =
		\exp[-i\chi(t)\sqrt{\omega^2-\lambda D^2}] c_t.
\end{equation}
For the special case in which $\chi$ is a {\em constant\/}, the right
hand side corresponds to a unitary transformation on the
`one-particle' space $L^2(\mathR,dt)$, and hence the time-averaged
energy $\int_{-\infty}^\infty dt\,H_t$ exists as a genuine operator
on the associated Fock space. Of course, the eigenvalues and
eigenvectors of this operator are well-known for the quantum field
theory analogue of the total Hamiltonian operator $\int dx\, H(x)$;
in the history theory, the associated projection operators
correspond to appropriate propositions about the value of the time
average of the energy.

	Note that the relation between $x_t$ and the new annihilation
and creation operators is non-local in time:
\begin{equation}
	x_t=\left({\hbar\over 2m\sqrt{\omega^2-\lambda
			D^2}}\right)^{1/2}(c_t+c_t^\dagger).
\end{equation}
In particular, the correlation function is 
\begin{equation}
	\bra{0}x_tx_s\ket{0}= 
	\left({\hbar\over 2m\sqrt{\omega^2-\lambda D^2}}\right)
		\delta(t-s)
\end{equation}
where the non-local quantity on the right hand side is the Green's
function of the elliptic, partial differential operator
$\sqrt{\omega^2-\lambda D^2}$. Thus we still have a Gaussian
stochastic process but it is `softer' than the one constructed
earlier whose correlation function was Eq.\ (\ref{<xtxs>}). Of
course, existence of non-local terms is a common occurrence in
normal relativistic quantum field theory, but there the non-locality
is in space, not time. It remains to be seen whether the
`velocity-extended' Hamiltonian in Eq.\ (\ref{Def:Hvext}) has any
real physical application in the consistent histories theory.

\section{Quantum Field Theory}
\subsection{The canonical history algebra} 
We wish now to extend the discussion to the HPO theory of a free
scalar field. Hartle \cite{Har93a} proposed a consistent histories
approach to quantum field theory based on path integrals, and
Blencowe \cite{Ble91} gave a careful analysis of the use of class
operators.  However, almost nothing has been said about the HPO
scheme in this context, and we shall now briefly present the
necessary developments.  The resemblance of the history version of
quantum mechanics (`field theory in zero spatial dimensions') to a
canonical field theory in one spatial dimension suggests that the
history version of quantum field theory in three spatial dimensions
should resemble canonical quantum field theory in {\em four\/}
spatial dimensions.  We shall see that this expectation is fully
justified.

	The first step in constructing an HPO version of quantum field
theory is to foliate four-dimensional Minkowski space-time with the
aid of a time-like vector $n^\mu$ that is normalised by
$\eta_{\mu\nu}n^\mu n^\nu=1$, where the signature of the Minkowski
metric $\eta_{\mu\nu}$ has been chosen as $(+,-,-,-)$. The canonical
commutation relations for a standard bosonic quantum field theory
(the analogue of Eq.\ (\ref{CCR})) in three spatial dimensions are
\begin{eqnarray}
	&&{[\,}\phi(\underline{x}_1),\,\phi(\underline{x}_2)\,]=0
													\label{[phiphi]} \\ 
	&&{[\,}\pi(\underline{x}_1),\,\pi(\underline{x}_2)\,]=0 
													\label{[pipi]} \\
	&&{[\,}\phi(\underline{x}_1),\,\pi(\underline{x}_2)\,]=
i\hbar\delta^3(\underline{x}_1-\underline{x}_2) \label{[phipi]}
\end{eqnarray}
where $\underline{x}_1$ and $\underline{x}_2$ are three-vectors
that are spatial with respect to the foliation vector $n$. In
constructing the associated HPO theory we shall assume that the
passage from the canonical algebra Eq.\ (\ref{CCR}) to the history
algebra Eqs.\ (\ref{ctsHWxx})--(\ref{ctsHWxp}) is reflected in the
field theory case by passing from Eqs.\
(\ref{[phiphi]})--(\ref{[phipi]}) to
\begin{eqnarray}
&&{[\,}\phi_{t_1}(\underline{x}_1),\,\phi_{t_2}(\underline{x}_2)\,]=0
										\label{[tphiphi]}		\\
	&&{[\,}\pi_{t_1}(\underline{x}_1),\,\pi_{t_2}(\underline{x}_2)\,]=0
										\label{[tpipi]}			\\
	&&{[\,}\phi_{t_1}(\underline{x}_1),\,\pi_{t_2}(\underline{x}_2)\,]=
		i\hbar\delta(t_1-t_2)\delta^3(\underline{x}_1-\underline{x}_2)
										\label{[tphipi]}
\end{eqnarray}
where, for each $t\in\mathR$, the fields $\phi_t(\underline{x})$ and
$\pi_t(\underline{x})$ are associated with the spacelike
hypersurface $(n,t)$ whose normal vector is $n$ and whose foliation
parameter is $t$; in particular, the three-vector $\underline{x}$ in
$\phi_t(\underline{x})$ or $\pi_t(\underline{x})$ denotes a vector
in this space.

	In using this algebra, we have in mind a representation that is
some type of continuous tensor product $\otimes_{t\in\mathR}{\cal
H}_t$ where each ${\cal H}_t$ carries a representation of the
standard canonical commutation relations Eqs.\
(\ref{[phiphi]})--(\ref{[phipi]}) for a scalar field theory
associated with the given spacetime foliation. However, to emphasise
the underlying spacetime picture it is convenient to rewrite Eqs.\
(\ref{[tphiphi]})--(\ref{[tphipi]}) in terms of four-vectors $X$ and
$Y$ as
\begin{eqnarray}
	&&{[\,}\phi(X),\,\phi(Y)\,]=0				\label{[phiXphiY]}\\
	&&{[\,}\pi(X),\,\pi(Y)\,]=0					\label{[piXpiY]}  \\
	&&{[\,}\phi(X),\,\pi(Y)\,]=i\hbar\delta^4(X-Y).\label{[phiXpiY]} 
\end{eqnarray}
In relating these expressions to those in Eqs.\
(\ref{[tphiphi]})--(\ref{[tphipi]}) the three-vector $\underline{x}$
may be equated with a four-vector $x_n$ that satisfies $n\cdot
x_n=0$ (the dot product is taken with respect to the Minkowski
metric $\eta_{\mu\nu}$) so that the pair
$(t,\underline{x})\in\mathR\times\mathR^3$ is associated with the
spacetime point $X=tn+x_n$ (in particular, $t=n\cdot X$). Note,
however, that the covariant-looking nature of these expressions is
deceptive and it is not correct to assume {\em a priori\/} that the
fields $\phi(X)$ and $\pi(Y)$ transform as spacetime scalars under
the action of some `external' spacetime Poincar\'e group that acts
on the $X$ and $Y$ labels---as things stand there is an implicit $n$
label on both $\phi$ and $\pi$. We shall return to this question
later.

\subsection{The Hamiltonian Algebra}	
The key idea of our HPO approach to quantum field theory is that the
physically-relevant representation of the canonical history algebra
Eqs.\ (\ref{[tphiphi]})--(\ref{[tphipi]}) [or, equivalently, Eqs.\
(\ref{[phiXphiY]})--(\ref{[phiXpiY]})] is to be selected by
requiring the existence of operators that represent history
propositions about temporal averages of the energy defined with
respect to the chosen spacetime foliation. Thus, for a fixed
foliation vector $n$, we seek a family of `internal' Hamiltonians
$H_{n,t}$, $t\in\mathR$, whose explicit formal form ({\em i.e.}, the
analogue of Eq.\ (\ref{Def:Ht})) can be deduced from the standard
quantum field theory expression to be
\begin{equation}
	H_{n,t}:={1\over 2}\int d^4X\left\{\pi(X)^2+
	(n^\mu n^\nu-\eta^{\mu\nu})\partial_\mu\phi(X)\partial_\nu\phi(X)
		+m^2\phi(X)^2\right\}\delta(t-n\cdot X).\label{Def:Hnt} 
\end{equation}
The analogous, temporally-averaged object is 
\begin{eqnarray}
	H_n(\chi)&:=&\int_{-\infty}^\infty dt\,\chi(t) H_{n,t}	
												\label{Def:Hnchi}\\
		&=&{1\over 2}\int d^4X\left\{\pi(X)^2+
	(n^\mu n^\nu-\eta^{\mu\nu})\partial_\mu\phi(X)\partial_\nu\phi(X)
		+m^2\phi(X)^2\right\}\chi(n\cdot X)				\nonumber 
\end{eqnarray}
where $\chi$ is a real-valued test function.

	As in the discussion above of the simple harmonic oscillator,
the next step is to consider the commutator algebra that would be
satisfied by the operators $H_n(\chi)$ {\em if\/} they existed.
These field-theoretic analogues of Eqs.\
(\ref{[Hchixt]})--(\ref{[Hchi1Hchi2]}) are readily computed as
\begin{eqnarray} 
	&&{[\,}H_n(\chi),\,\phi(X)\,]=-i\hbar\chi(n\cdot X)\pi(X) 
													\label{[HchiphiX]}\\
	&&{[\,}H_n(\chi),\,\pi(X)\,]=i\hbar\chi(n\cdot X) K_n\phi(X)
													\label{[HchipiX]}\\ 
	&&{[\,}H_n(\chi_1),\,H_n(\chi_2)\,]=0
\end{eqnarray} 
where $K_n$ denotes the partial differential operator
\begin{equation} 
(K_nf)(X):=\left[(\eta^{\mu\nu}-n^\mu n^\nu)
\partial_\mu\partial_\nu +m^2 \right]f(X). \label{Def:Kn}
\end{equation}

	The exponentiated form of Eqs.\
(\ref{[HchiphiX]})--(\ref{[HchipiX]}) is 
\begin{eqnarray}
	e^{iH_n(\chi)/\hbar}\,\phi(X)\,e^{-iH_n(\chi)/\hbar}&=&
	\cos\left[\chi(n\cdot X)\sqrt{K_n}\right]\phi(X)
		+{1\over\sqrt K_n}\sin\left[\chi(n\cdot X)
		\sqrt{K_n}\right]\pi(X)				\label{AutphiX}	\\
	e^{iH_n(\chi)/\hbar}\,\pi(X)\,e^{-iH_n(\chi)/\hbar}&=&
	-\sqrt{K_n}\sin\left[\chi(n\cdot X)\sqrt{K_n}\right]\phi(X)
+\cos\left[\chi(n\cdot X)\sqrt{K_n}\right]\pi(X)	\label{AutpiX} 
\end{eqnarray}
where the square-root operator $\sqrt{K_n}$, and functions thereof,
can be defined rigorously using the spectral theory of the
self-adjoint, partial differential operator $K_n$ on the Hilbert
space $L^2(\mathR^4,d^4X)$. Note that the expression $\chi(n\cdot
X)\sqrt{K_n}$ is unambiguous since, viewed as an operator on
$L^2(\mathR^4,d^4X)$, multiplication by $\chi(n\cdot X)$ commutes
with $K_n$.

\subsection{The Fock Space Representation} 
The right hand side of Eqs.\ (\ref{AutphiX})--(\ref{AutpiX}) defines
an automorphism of the CHA Eqs.\
(\ref{[phiXphiY]})--(\ref{[phiXpiY]}) and the task is to find a
representation of the latter in which these automorphisms are
unitarily implemented. To this end, define new operators
\begin{eqnarray}
q(X)&:=&K^{1/4}_n\phi(X) \label{Def:qX}\\ p(X)&:=&K^{-1/4}_n\pi(X)
\label{Def:pX} 
\end{eqnarray} and
\begin{equation}	
	b(X):={1\over\sqrt2}\Big(q(X)+ip(X)\Big)=
	  {1\over\sqrt2}\Big(K^{1/4}_n\phi(X)+iK^{-1/4}_n\pi(X)\Big)
												\label{Def:bX}
\end{equation}
which satisfy
\begin{eqnarray}
		&&{[\,}b(X),\,b(Y)\,]=0					\\
		&&{[\,}b^\dagger(X),\,b^\dagger(Y)\,]=0		\\
		&&{[\,}b(X),\,b^\dagger(Y)]=\hbar\delta^4(X-Y).
\end{eqnarray}

Then
\begin{eqnarray}
	e^{iH_n(\chi)/\hbar}\,q(X)\,e^{-iH_n(\chi)/\hbar}&=&
	\cos\left[\chi(n\cdot X)\sqrt{K_n}\right]q(X)
			+\sin\left[\chi(n\cdot X)\sqrt{K_n}\right]p(X)\\
	e^{iH_n(\chi)/\hbar}\,p(X)\,e^{-iH_n(\chi)/\hbar}&=&
		-\sin\left[\chi(n\cdot X)\sqrt{K_n}\right]q(X)
	  		+\cos\left[\chi(n\cdot X)\sqrt{K_n}\right]p(X) 
\end{eqnarray}
and so
\begin{equation}
	e^{iH_n(\chi)/\hbar}\,b(X)\,e^{-iH_n(\chi)/\hbar}=
		e^{-i\chi(n\cdot X)\sqrt{K_n}}\,b(X).	\label{Autb(X)} 
\end{equation}
However, the operator defined on $L^2(\mathR^4)$ by 
\begin{equation}
	(U(\chi)\psi)(X):=e^{-i\chi(n\cdot X)\sqrt{K_n}}\psi(X) 
\end{equation}	
is unitary, and hence---using the same type of argument invoked
earlier for the simple harmonic oscillator---we conclude that the
desired quantities $H_n(\chi)$ exist as self-adjoint operators on
the Fock space ${\cal F}[L^2(\mathR^4,d^4X)]$ associated with the
creation and annihilation operators $b^\dagger(X)$ and $b(X)$. The
spectral projectors of these operators then represent propositions
about the time-averaged value of the energy in the spacetime
foliation determined by $n$.

\subsection{The Question of External Lorentz Invariance} 
An important part of standard quantum field theory is a proof of
invariance under the Poincar\'e group---something that, in the
canonical formalism, is not totally trivial since the
Schr\"odinger-picture fields depend on the reference frame ({\em
i.e.}, the spacetime foliation). The key ingredient is a
construction of the generators of the Poincar\'e group as explicit
functions of the canonical field variables; in practice, the first
step is often to construct the Heisenberg-picture fields with the
aid of the Hamiltonian, and then to demonstrate manifest Poincar\'e
covariance within that framework.  The canonical fields associated
with any spacelike surface in a particular Lorentz frame can then be
obtained by restricting the Heisenberg fields (and their normal
derivatives) to the surface.

	When considering the role of the Poincar\'e group in the HPO
picture of consistent histories, the starting point is the
observation that, heuristically speaking, for a given foliation
vector $n$---and for each value of the associated time $t$---there
will be a Hilbert space ${\cal H}_t$ carrying an independent copy of
the standard quantum field theory. In particular, therefore, for
fixed $n$, there will be a representation of the Poincar\'e group
associated with each spacelike slice $(n,t)$, $t\in\mathR$.  Thus if
$A_a$, $a=1,2,\ldots,10$ denote the generators of the Poincar\'e
group, there should exist a family of operators $A_t^a$ which, for
each $t\in\mathR$, generate the `{\em internal\/}' Poincar\'e group
${\cal P}_{n,t}$ associated with the slice $(n,t)$.  These operators
will satisfy a `temporally gauged' version of the Poincar\'e
algebra.  More precisely, if $C^{ab}{}_c$ are the structure
constants of the Poincar\'e group, so that
\begin{equation}
	[\,A^a,\,A^b\,]=iC^{ab}{}_cA^c,
\end{equation}
then the algebra satisfied by the history theory operators $A_t^a$
is
\begin{equation}
	[\,A^a_t,\,A^b_s\,]=i\delta(t-s) C^{ab}{}_cA_t^c
\end{equation}
which, of course, reflects the way in which the canonical
commutation relations Eqs.\ (\ref{[phiphi]}--\ref{[phipi]}) are
replaced by Eqs.\ (\ref{[tphiphi]}--\ref{[tphipi]}) in the history
theory.

	As always in quantum theory, the energy operator is of
particular importance, and in the present case we have a family of
Hamiltonian operators $H_{n,t}$, $t\in\mathR$, which are related to
the generators $P_{n,t}^\mu$ of translations for the quantum field
theory associated with the hypersurface $(n,t)$ by
\begin{equation}
	H_{n,t}=n_\mu P_{n,t}^\mu.
\end{equation}
In fact, it is straightforward to show that
\begin{equation}
	P_{n,t}^\mu=n^\mu H_{n,t}+\int d^4X\,\delta(t-n\cdot X)
		(n^\mu n\cdot\partial\phi-\partial^\mu\phi)\pi
\end{equation}
which suggests that, as would be expected, the components of
$P_{n,t}^\mu$ normal to $n$ act are the generators of spatial
translations in the hypersurface $(n,t)$. Indeed, Eq.\
(\ref{[HchiphiX]}) generalises to
\begin{equation}
	{[\,}P_n^\mu(\chi),\,\phi(X)\,]=-i\hbar\chi(n\cdot X)
	\big\{n_\mu\pi(X)+(\partial_\mu\phi(X)-
		n_\mu\, n\cdot\partial\phi(X))\big\}. 
\end{equation}
Similarly, the `temporally gauged' Lorentz generators satisfy
\begin{eqnarray}
\lefteqn{{[\,}J_{n,t}^{\mu\nu},\,\phi(X)\,] = }			\\
		& & \qquad i\hbar\delta(t-n\cdot X)
		\big\{X^\mu(\partial^\nu\phi-n^\nu n\cdot\partial\phi)
			- X^\nu(\partial^\mu\phi-n^\mu n\cdot\partial\phi)
			-(X^\mu n^\nu-X^\nu n^\mu)\pi\big\}\nonumber.
\end{eqnarray}

	As emphasised above, each generator of the group ${\cal
P}_{n,t}$ acts `internally' in the Hilbert space ${\cal H}_t$; in
particular, this is true of the Hamiltonian, which (modulo the need
to smear in $t$) generates translations along an `internal' time
label $s$ that is to be associated with each leaf $(n,t)$ of the
foliation. It is important to note that $H_{n,t}$ does {\em not\/}
generate translations along the `exernal' time parameter $t$ that
appears in the CHA Eqs.\ (\ref{[tphiphi]}--\ref{[tphipi]}) and which
labels the spacelike surface (of course, there is an analogous
statement for the Hamiltonians $H_t$ in the HPO model of the simple
harmonic oscillator considered earlier). The existence of these
internal Poincar\'e groups is sufficient to guarantee covariance of
physical quantities, such as transition amplitudes, that can be
calculated in the class operator version of the theory.

	However, the HPO formalism admits an additional type of
Poincar\'e group---what we shall call the `{\em external\/}'
Poincar\'e group---which is defined to act on the pair of labels
$(\underline{x},t)$ that appear in the CHA Eqs.\
(\ref{[tphiphi]}--\ref{[tphipi]}). Thus these labels include the
`external' time parameter $t$ that specifies the leaf $(n,t)$ of the
foliation associated with the timelike vector $n$. In the context of
the covariant-looking version Eqs.\
(\ref{[phiXphiY]}--\ref{[phiXpiY]}) of the CHA, the main question is
whether the fields $\phi(X)$ and $\pi(X)$ transform in a covariant
way under this external group.

	As far as the field $\phi(X)$ is concerned it seems reasonable
to consider the possibility that this may an external scalar in the
sense that there exists a unitary representation $U(\Lambda)$ of the
external Lorentz group $U(\Lambda)$ such that
\begin{equation}
	U(\Lambda)\phi(X)U(\Lambda)^{-1}=\phi(\Lambda X).
\end{equation}
The spectral projectors of the (suitably smeared) operators
$\phi(X)$ then represent propositions about the values of the
spacetime field in a covariant way. 

	However, the situation for the
field momentum $\pi(X)$ is different since this is intrinsically
associated with the timelike vector $n$.  Indeed, the natural thing
would be to require the existence of a {\em family\/} of operators
$\pi_n(X)$ where $n$ lies in the hyperboloid of all timelike
(future-pointing) vectors, and such that
\begin{equation}
		U(\Lambda)\pi_n(X)U(\Lambda)^{-1}=
					\pi_{\Lambda n}(\Lambda X).
\end{equation}
The next step in demonstrating external Poincar\'e covariance would
be to extend the algebra (\ref{[phiXphiY]}--\ref{[phiXpiY]}) to
include the $n$ parameter on the $\pi$ field; in particular, one
would need to specify the commutator $[\,\pi_n(X),\,\pi_m(Y)\,]$,
but it is not obvious {\em a priori} what this should be.

	Another possibility would be to try to combine the Heisenberg
picture---and its associated `internal' time $s$---with the external
time parameter $t$ of the spacetime foliation to give some scheme
that was manifestly covariant in the context of a five-dimensional
space with signature $(+++,--)$ associated with the variables
$(\underline{x},t,s)$. However, we do not know if this is possible and
the demonstration of external Poincar\'e covariance, if it exists,
remains the subject for future research.

\section{Conclusions}
We have discussed the introduction of continuous-time histories
within the `HPO' version of the consistent-histories formalism in
which propositions about histories of the system are represented by
projection operators on a `history' Hilbert space. The history
algebra (whose representations specify this space) for a particle
moving in one dimension is isomorphic to the canonical commutation
relations for a one-dimensional quantum field theory, thus
allowing the history theory to be studied using techniques drawn from
quantum field theory. In particular, we have shown how the problem
of the existence of infinitely many inequivalent representations of
the history algebra can be solved by requiring the existence of
operators whose spectral projectors represent propositions about
time-averages of the energy.

	We have shown how the Heisenberg picture is changed in the HPO
formalism in such a way that the familiar, partial-differential
equations of motion are replaced by functional differential
equations; these operators are used in the proof that the angular
momentum operators of the three-dimensional theory are anomaly free.
The question of potential anomalies in the history algebra is rather
intriguing and it would be interesting to study a theory in which
such things might be expected, such as a fermionic system.  

	A striking property of the HPO formalism is the potential
existence of a velocity operator that commutes with the position
operators, thus opening up a new perspective on the old debate about
the operational meaning of the Heisenberg uncertainty relations
between position and momentum. The introduction of the velocity
operator suggests a number of topics for future work: for example,
it would be most interesting to see if some thing like the action
functional of the classical action has a natural role to play in the
HPO theory.

	Finally, we have shown how the HPO scheme can be extended to
the history version of canonical quantum field theory. We discussed
the difference between the `internal' and `external' Poincar\'e
groups and indicated how the former are implemented in the formalism.
A major challenge for future research is to construct an HPO quantum
field theory which is manifestly covariant under this external
symmetry group.

\bigskip\noindent
{\bf Acknowledgements}

Noah Linden is grateful to the Leverhulme and Newton Trusts for
their continued support. Dina Savvidou thanks the NATO Science
Fellowship Programme for their support.

\end{document}